\documentclass[
  aps,
  twocolumn,
  superscriptaddress,
  amsmath,amssymb,
  floatfix,
  nobibnotes,
  nofootinbib
]{revtex4-2}
\setcitestyle{super}

\usepackage{amsmath}
\usepackage{amssymb}
\usepackage{bm}
\usepackage{upgreek}
\usepackage{fdsymbol}
\usepackage{graphicx}
\usepackage{booktabs}
\usepackage{multirow}
\usepackage{array}
\usepackage{xcolor}
\usepackage[colorlinks=true,allcolors=blue]{hyperref}
\usepackage{textcomp}
\usepackage{lettrine}

\makeatletter
\renewcommand\frontmatter@abstractwidth{\dimexpr\textwidth-0.5in\relax}
\makeatother

\usepackage{fancyhdr}
\usepackage{titlesec}

\titleformat{\section}
  {\normalfont\small\bfseries\raggedright}
  {\thesection}
  {1em}
  {}

\titleformat{\subsection}[runin]
  {\normalfont\small\bfseries\raggedright}
  {\thesubsection}
  {1em}
  {}[.]

\titlespacing*{\section}
  {0pt}{\baselineskip}{0pt}

\titlespacing*{\subsection}
  {0pt}{0pt}{0.5em}

\renewcommand{\selectlanguage}[1]{}

\begin{document}

\title{\LARGE 
Direct visualization of charge dynamics by Pockels imaging \\in optoelectronics}

\author{Adriano~Cola$^*$}
\affiliation{CNR IMM, Institute for Microelectronics and Microsystems,
Via Monteroni, 73100 Lecce, Italy}

\author{Antonio~Valletta}
\affiliation{CNR IMM, Institute for Microelectronics and Microsystems,
Via del Fosso del Cavaliere 100, 00133 Roma, Italy}

\author{Isabella~Farella}
\affiliation{CNR IMM, Institute for Microelectronics and Microsystems,
Via Monteroni, 73100 Lecce, Italy}

\author{Vaclav~Dedi\v{c}}
\affiliation{Faculty of Mathematics and Physics, Charles University,
Ke Karlovu 5, 121 16 Prague, Czech Republic}

\author{Lorenzo~Dominici$^*$,}
\affiliation{CNR NANOTEC, Institute of Nanotechnology,
Via Monteroni, 73100 Lecce, Italy\\
$^*$email \href{mailto:adriano.cola@cnr.it}{adriano.cola@cnr.it} , 
\href{mailto:lorenzo.dominici@cnr.it}{lorenzo.dominici@cnr.it} \\
Published version at Matter \& Light 1, 100141, 2026 \\ 
\href{https://doi.org/10.1016/j.matlit.2026.100141}{https://doi.org/10.1016/j.matlit.2026.100141}
}


\begin{abstract}
\noindent \textbf{Understanding charge dynamics in optoelectronic
devices is essential because it governs carrier transport, signal
evolution, and device performance. Yet, experimental access
to the spatiotemporal evolution of charge remains limited. Here, we introduce
Carrier Dynamics by Pockels Imaging (CDPI), a sensitive, quantitative,
and non-invasive pump-probe technique based on the linear
electro-optic (Pockels) effect. CDPI images the 
electric-field wave associated with traveling photogenerated carriers
and reveals the evolution of charge clouds in space and
time. This provides direct access to fundamental transport processes,
including drift, diffusion, carrier-carrier interactions, trapping,
and detrapping. Furthermore, the internal conduction- and
displacement-current profiles are obtained from the electric-field
dynamics and shown to be consistent with the externally measured
current transients. Demonstrated in CdTe radiation detectors with
micrometre-scale spatial resolution and nanosecond temporal resolution,
CDPI is applicable to materials exhibiting a measurable Pockels effect,
including electro-optic semiconductors, oxides, halide perovskites,
organic materials, and emerging low-dimensional systems,
providing a platform to investigate charge transport
across optoelectronics.}
\end{abstract}


\maketitle



\lettrine{E}{lectronic} and optoelectronic devices operate through charge transport,
whose microscopic dynamics ultimately determine their functionality.
External and built-in electric fields define the pathways followed by
charge carriers, while the carriers themselves continuously reshape the
internal electric-field distribution through their associated space
charge. This mutual interplay can be pictured as heavy marbles locally
deforming a soft inclined slipway: the carriers move along the field
landscape while simultaneously modifying it. Imaging this evolving
electric field therefore provides direct access to charge dynamics,
revealing the link between carrier transport and the electrical
response of the device.

Electric-field-induced modulation of optical properties, particularly
of the refractive index, provides a powerful, quantitative, and
non-invasive means of probing local electric fields. Among
electro-optic effects, the linear Pockels effect, exhibited by
noncentrosymmetric materials, is particularly attractive because its
response is fundamentally limited only by the electronic polarizability
of the medium, enabling ultrafast operation. These intrinsic
capabilities have made the Pockels effect a key building block for
high-speed optical modulation
\cite{abel_large_2019,alexander_nanophotonic_2018,powell_integrated_2022,thomaschewski_pockels_2022}
and electro-optic sampling in integrated photonics and electronic
diagnostics
\cite{kolner_electrooptic_1986,donaldson_electro-optic_1990,peterkin_electric_1995,ito_high-speed_2004,currie_performance_2013}.

The Pockels effect has progressively evolved into an established tool
for mapping electric-field distributions inside semiconductor radiation
detectors
\cite{sellin_electric_2010,cola_investigation_2006,dedic_mapping_2021}.
Early studies clarified the microscopic origin of bias-induced
polarization \cite{cola_polarization_2009} by reconstructing
electric-field distributions under dark operating conditions. More
recently, attention shifted towards radiation-induced polarization,
aiming to understand and control the electric-field redistribution
produced by optical excitation
\cite{cola_cdte_2014,cola_electric-field_2023,cola_optical_2022,franc_control_2015}.
These investigations, including our recent studies on two-dimensional
field mapping \cite{cola_electric-field_2023} and optical doping
\cite{cola_optical_2022}, remained confined to stationary or
quasi-stationary regimes, where the observed field evolution is mainly
governed by trapped space charge. Consequently, despite the
intrinsically ultrafast nature of the Pockels response, its potential
for directly probing the dynamics of photogenerated free carriers has
remained largely unexplored.

Beyond Pockels-based approaches, carrier and electric-field dynamics
have been investigated through several techniques, including other
nonlinear optical effects such as the Kerr effect
\cite{sima_time-continuous_2015} and electric-field-induced
second-harmonic generation
\cite{bhattacharya_nonlinear_2023,shyam_direct_2025,cao_electric_2021,yin_win_transient_2019,manaka_direct_2007},
as well as transient absorption \cite{gabriel_direct_2013},
photoluminescence imaging
\cite{sridharan_time-resolved_2019,li_ultrafast_2023}, and various
ultrafast microscopies
\cite{terada_real-space_2010,sun_mapping_2016,wang_real-space_2025,wang_crystallographic_2023,man_imaging_2017,son_imaging_2014,sirotin_direct_2023}.
These methods have enabled remarkable spatial and temporal resolutions
and have provided important insights into carrier generation,
diffusion, recombination, and related processes (for a review on
spatiotemporal imaging of photogenerated charge carriers, see Wang et al.~\cite{wang_spatiotemporal_2025}). However, they generally probe
different physical observables, are often perturbative, and provide
access only to the surface or to specific aspects of carrier dynamics,
rather than directly following the coupled evolution of carrier
distributions and the electric field. An extended comparison of
existing free-carrier and electric-field imaging techniques is reported
in Table~S1.

In this work, we introduce Carrier Dynamics by Pockels Imaging (CDPI),
a pump-probe electro-optic imaging technique that extends
Pockels-based field mapping from quasi-static space-charge studies to
the nanosecond dynamics of photogenerated free carriers. CDPI images
the electric-field perturbation generated by photogenerated carriers
and, through Maxwell's equations together with the standard
drift-current relation, reconstructs the spatiotemporal evolution of
electron clouds together with the local conduction current and the
displacement current distributed throughout the device. This
establishes a unified framework for investigating charge transport and
its electrical response in electro-optic materials and optoelectronic
systems.

We validate CDPI using bulky CdTe radiation detectors, investigating
electron dynamics with nanosecond temporal and micrometre-scale spatial
resolution as a function of applied bias and optical excitation. By
progressively increasing the photogenerated charge density, we also
explore highly perturbative transport regimes that have previously been
investigated almost exclusively through externally measured current
transients. The experiments are complemented by numerical simulations
closely reproducing the experimental protocol. Together, the
experiments and simulations demonstrate the ability of CDPI to directly
visualize charge dynamics and relate them to the underlying
electric-field evolution. Although the present implementation is
demonstrated in bulk CdTe devices, the intrinsic bandwidth of the
Pockels effect combined with spatially resolved electric-field imaging
provides a pathway towards investigating substantially faster carrier
dynamics in a broad range of electro-optic materials and
optoelectronic devices.

\begin{figure*}[t]
  \includegraphics[width=0.9\textwidth]{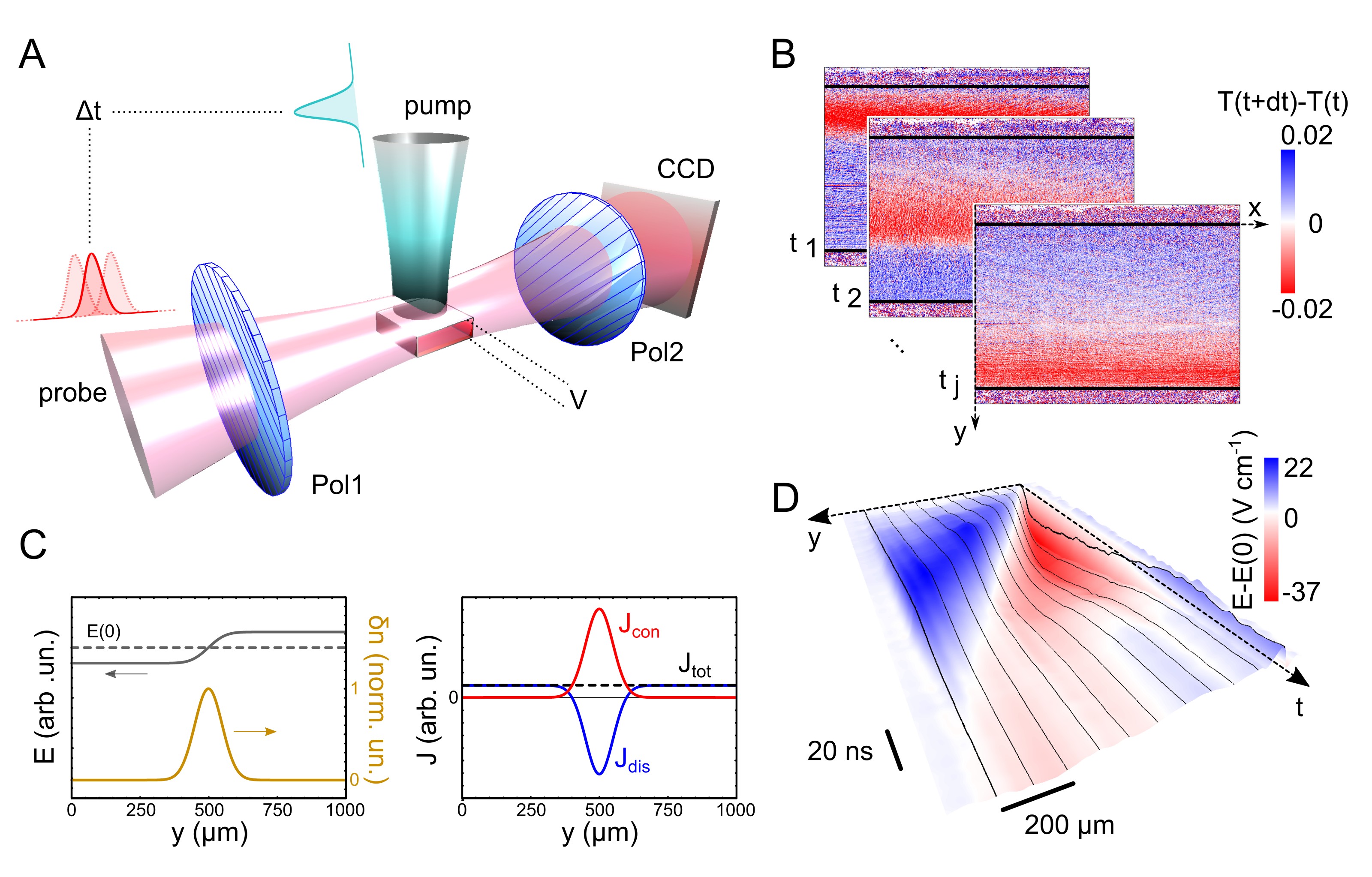}
  \caption{\label{fig:1}%
    \textbf{Experimental setup and working principles.}\\
    (A)~Scheme of the working configuration. Two independent laser
    pulses are temporally shifted by means of an electronic delay and
    sent on the device to work as a probe and a pump beam. The device
    is kept at a given voltage $V$ and placed between two crossed
    linear polarizers (Pol1 and Pol2). The probe beam crossing the
    sample and polarizers is imaged on a CCD camera.
    (B)~Sequence of processed real-space images with the absolute
    transmittance variations at different time frames
    ($t_1, t_2, \dots, t_j$).
    (C)~Representative profiles of electric field and charge cloud
    (left subpanel) and conduction, displacement and total current
    (right subpanel). The cathode and anode are placed at $y=0$ and
    $y=1000$ $\mu$m, respectively.
    (D)~Experimental time-space chart of the electric-field difference
    with respect to initial (just before the pump pulse) time. The
    applied voltage is 50 V, and the time range is 280 ns. Spatial and
    temporal scale bars represent 200 $\mu$m and 20 ns, respectively.%
  }
\end{figure*}

\section*{Carrier Dynamics by Pockels Imaging (CDPI)}

The initial outcome of the CDPI experiments is a series of transmittance
Pockels images $T(x,y,t)$, each one corresponding to a progressive
delay of the probing pulse with respect to the excitation pulse (see
Fig.~\ref{fig:1}A for the experimental scheme and Methods for details).

The excitation pulse creates a thin sheet of electron-hole pairs just
below the irradiated semi-transparent contact, which moves under the
applied electric field. In the present configuration, the irradiated
contact is the cathode, which implies that holes are quickly collected
while electrons move towards the bottom anode. Hence, the dynamics
that we address here refer to electrons, but the following analysis
can also be applied to holes in case of anode irradiation.

A preliminary visualization of the photogenerated electron sheet is
shown in Fig.~\ref{fig:1}B. The panel contains differential
transmittance images $T(x,y,t+\mathrm{d}t)-T(x,y,t)$ at selected
instants, where transmission variations of a few percent are resolved.
The motion of the charge distribution is revealed by the intensity
strip (red color) moving downward to the collecting electrode (Video
S1). Given the planar geometry of the present device, we focus on the
1D profiles (along the $y$-direction) and their dynamics, by averaging
along the $x$-direction.

Upon pump excitation, there is always a propagating (usually
sigmoid-like) wave in the electric field associated with the moving
(usually Gaussian) carrier cloud of concentration $\delta n(y,t)$, as
schematically shown in Fig.~\ref{fig:1}C (left panel). This propagating
wave superimposes to the background electric field, the pre-existent
field profile $E(t=0)$ under dark which is assumed as constant in the
scheme. In general, such a field wave can be a small or large
perturbation on a non-uniform spatial field profile. The CDPI technique
offers a large operative dynamic range and in particular the capability
to unveil minimal field perturbations $\Delta E$ of the order of a few
V cm$^{-1}$.

In terms of currents, the conduction current is localized as a peak on
$\delta n(y,t)$, while the displacement current exhibits a spatially
extended contribution together with an opposite-sign peak associated
with the moving carrier cloud. This behavior is depicted in
Fig.~\ref{fig:1}C (right panel), where the two peaked contributions
(red and blue solid lines) represent the individual currents and the
flat sum is the total current $J(t)$ (dashed black line). The chart in
Fig.~\ref{fig:1}D shows how the CDPI technique can provide powerful
insights. It illustrates the full time-space dynamics of field
variation $E(y,t)-E(y,t=0)$ at a glance. Such a chart is at the basis
for deriving the other involved quantities, including the displacement
current $\varepsilon\,\partial E(y,t)/\partial t$.

Hereafter we provide the main equations underlying CDPI. More details
are reported in Methods. From Gauss's law, the electron cloud
$\delta n(t,y)$ can be written in terms of the electric-field spatial
derivatives at the given and initial time, in this form:
\begin{equation}
\delta n(t,y)
  = \frac{\varepsilon}{q}
    \left[
      \frac{\partial E(y,t)}{\partial y}
      - \frac{\partial E(y,t=0)}{\partial y}
    \right],
\label{eq:1}
\end{equation}
where the second term accounts for the pre-existing charge distribution,
assumed stationary on the timescale of carrier motion.
Neglecting diffusion under the present operating conditions, the total
current $J(t)$, sum of conduction and displacement terms, can be thus
expressed as:
\begin{equation}
J(t)
  = \mu\,\varepsilon\,E(y,t)
    \left[
      \frac{\partial E(y,t)}{\partial y}
      - \frac{\partial E(y,t=0)}{\partial y}
    \right]
    + \varepsilon\,\frac{\partial E(y,t)}{\partial t},
\label{eq:2}
\end{equation}
where we note that $J(t)$ can be written only in terms of the electric
field, with the conduction and the displacement currents connected to
the spatial and temporal derivative of the electric field,
respectively. Hence Eq.~\eqref{eq:2} provides the central relation of
our approach: the experimentally measured electric-field evolution
enables the reconstruction of the spatial distributions of the conduction-
and displacement-current contributions. Their sum satisfies current
continuity and yields the total current flowing through the device,
which can be independently measured in the external circuit.
Equation~\eqref{eq:2} contains both spatial and temporal information
and provides two complementary routes to retrieve the carrier
distribution: directly from the spatial derivative of the electric
field through Eq.~\eqref{eq:1}, or alternatively from the
displacement-current contribution through the temporal derivative of
the electric field, according to:
\begin{equation}
\delta n_{\mathrm{displ}}(y,t)
  = -\,\frac{\varepsilon\,\partial E(y,t)/\partial t - J(t)}
           {q\,\mu\,E(y,t)}.
\label{eq:3}
\end{equation}

Together, Eqs.~\eqref{eq:1}--\eqref{eq:3} define the framework of CDPI,
where the experimentally measured electric-field evolution provides
access to charge distributions and local conduction- and
displacement-current contributions. Unlike conventional electrical
measurements, which generally provide only spatially integrated current
responses requiring auxiliary assumptions to infer internal transport
mechanisms, CDPI enables spatially resolved reconstruction of charge
and current dynamics.

Under the assumptions discussed in Methods, a global analytical
relationship between current and electric field can be obtained by
spatial integration of Eq.~\eqref{eq:2}, resulting in the following
approximate expression for the total current
\cite{weisz_electrode-limited_1968,batra_photocurrents_1970}:
\begin{equation}
J_{\mathrm{approx}}(t)
  = \frac{\varepsilon\,\mu}{2L}
    \left[
      E^{2}(L,t) - E^{2}(0,t)
    \right].
\label{eq:4}
\end{equation}
Compared with the local evaluation of Eq.~\eqref{eq:2}, which involves
spatial and temporal derivatives of the measured electric field and is
therefore more sensitive to experimental noise, Eq.~\eqref{eq:4} depends
only on the electric-field values at the electrodes. This makes it
particularly robust for the quantitative comparison with the externally
measured current transient. In summary, CDPI establishes a direct
correlation between the reconstructed internal carrier dynamics and the
total current flowing through the external circuit, providing a unified
view of microscopic charge transport and macroscopic device response.

\section*{Electron clouds at different voltages under low optical perturbation}
\begin{figure*}[t]
  \includegraphics[width=0.9\textwidth]{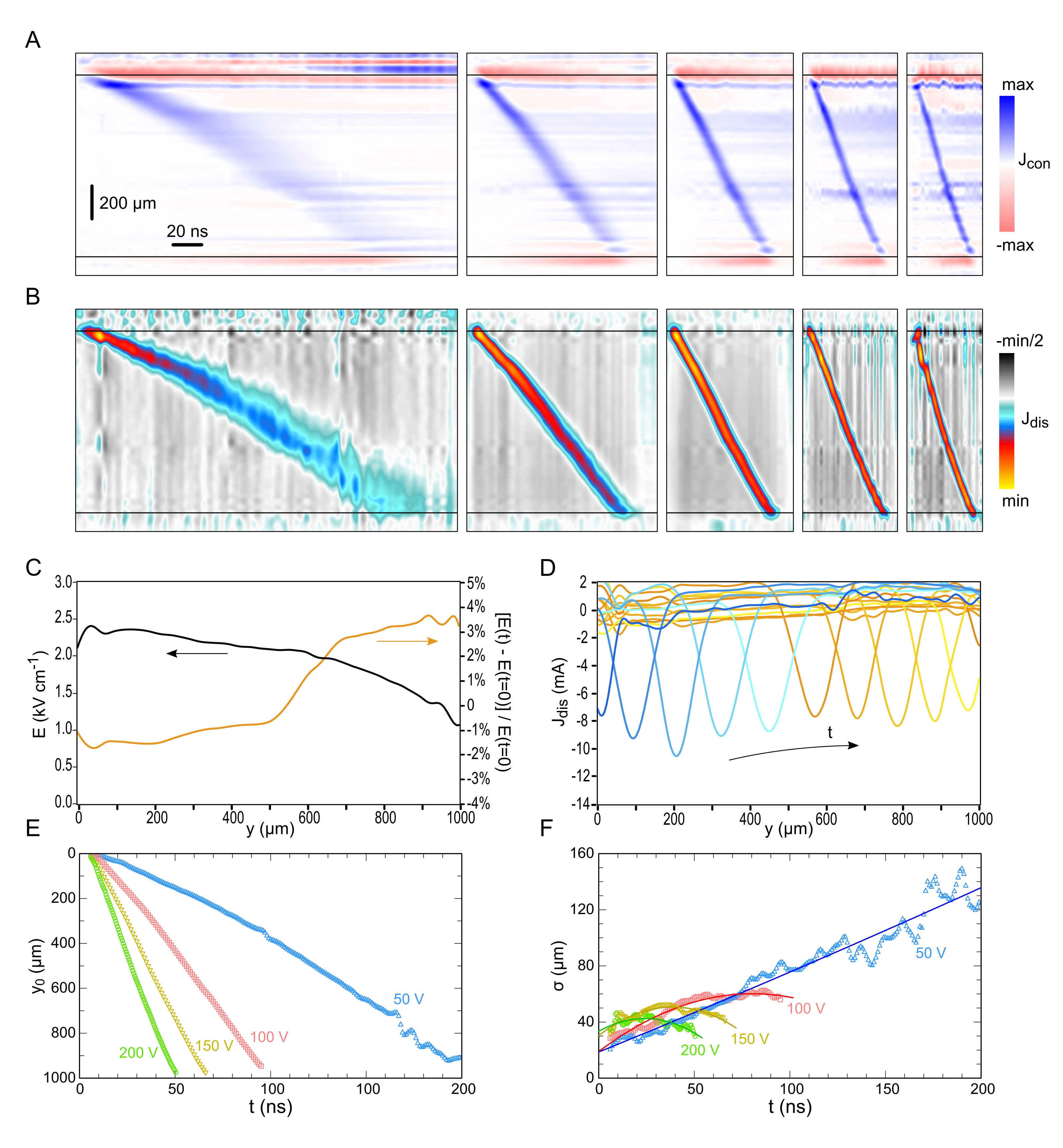}
  \caption{\label{fig:2}%
    \textbf{CDPI experiments at low power and different voltages.}\\
    (A)~Time-space charts of the conduction current evaluated using
    the spatial derivative of the electric field. The charts refer to
    the applied voltages of 50, 100, 150, 200, and 250~V, going
    left to right, respectively. The inner top and bottom horizontal
    solid lines represent the cathode ($y=0$) and anode
    ($y=1000$ $\mu$m) position, respectively. The minimum on the
    color bar has been kept fixed at minus the maximum value for the
    clarity and symmetry of representation (white areas correspond to
    a zero current). Spatial and temporal scale bars 
    represent 200 $\mu$m and 20 ns, respectively.
    (B)~Time-space charts of the displacement current evaluated from
    the temporal derivative of the electric field, for the same
    voltages of the previous charts. Here the maximum of the color bar
    has been kept fixed at minus half of the minimum value (white
    still corresponds to zero current). The time ranges are 240, 120,
    80, 60 and 48 ns going left to right, respectively, in (A) and (B).
    The time step is either 1 ns or 0.5 ns, and the time and space
    scale bars are the same through all the charts in (A) and (B).
    (C)~Electric field spatial profile (black solid line) and its
    relative variation with respect to the initial time (orange solid
    line). The profiles are taken at fixed time into the dynamics
    ($t=30$ ns).
    (D)~Spatial profiles of the displacement current $J_{\mathrm{dis}}(y)$ at
    time steps of 5 ns during the transit. Both (C) and (D) refer
    to the 200 V applied voltage.
    (E)~Peak position $y_0$ and
    (F)~cloud width $\sigma$ of the $J_{\mathrm{dis}}(y)$ clouds extracted from
    their Gaussian fit during time. The solid lines in (F) are a guide
    to the eye.%
  }
\end{figure*}

We first analyse the case of low optical perturbation, where the
electric field is only slightly perturbed by optical excitation
impinging on the cathode side of the device, biased at different
voltages. Due to the short penetration length (Fig.~S1), holes are
quickly collected while the electron cloud drifts towards the anode.
The clear motion of the electron cloud under different applied voltages
can be directly imaged. This is achieved by applying the spatial and
temporal derivative to the relevant time-space charts of the electric
field retrieved from EO experiments, like that of Fig.~\ref{fig:1}D.
The two associated quantities, the conduction and displacement
currents, are reported in Fig.~\ref{fig:2}A and Fig.~\ref{fig:2}B,
respectively. Their well-matched overlap confirms that they have been
correctly resolved and offer a good representation of the charge
carriers drifting from the cathode, where they are photogenerated, to
the anode ($y = 1000$ $\mu$m). It is possible to observe a sharp
localization of the electrons' cloud, moving with increasing speed at
larger applied voltages, while at the smallest voltage there is also an
appreciable spreading during the transit.

Looking in detail at the spatial features is possible, by reporting the
electric field at one given instant ($V = 200$ V, $t = 30$ ns), as
shown in Fig.~\ref{fig:2}C (black solid line). The profile appears to
be in very good agreement with the stationary ones measured on similar
devices in previous studies \cite{cola_cdte_2014,cola_relation_2016}.
Altogether, we report the relative variation of the electric field
(orange solid line) at that instant with respect to the initial time
(i.e. before the carrier cloud was photogenerated). An approximate
variation of less than 5\% can be noticed between the front and the
back of the cloud. This is a low perturbation condition, and the
electric field profile during the dynamics remains in fact not very
different from the one preceding the optical pulse (Video S2). The
associated displacement current $J_{\mathrm{displ}}(y)$ profiles are
shown in Fig.~\ref{fig:2}D at time intervals of 5 ns. The moving peak
at any time is smooth, symmetric and is well fitted by a Gaussian
profile (Video S3). We point out that these profiles can be assumed to
coincide almost exactly with the carriers' cloud. This association
stems from Eq.~\eqref{eq:3} due to the thin cloud's extension and has
been thoroughly verified. As a result, Figs.~\ref{fig:2}E and
\ref{fig:2}F show the temporal evolution of the center $y_0$ and width
$\sigma$ of the electron clouds extracted from the fit at different
$V$. Consistently, the collection time decreases as the voltage
increases. Furthermore, the slope of the $y_0(t)$ trajectories in
Fig.~\ref{fig:2}E is not linear, as it would be in a uniform electric
field. Instead, depending on the degree of non-uniformity of the field,
it exhibits either accelerating ($V = 50$ V) or decelerating concavity
($V > 100$ V). The cloud slowdown near the anode, for example, is due
to the electric field drop in that region (see Fig.~\ref{fig:2}C at
200 V).

Apart from speed, a more detailed analysis allows also to reveal that
the cloud dynamically changes its size at any of the applied voltages,
not only at the lowest one. In fact, while at $V = 50$ V the carrier
cloud continues to expand until the collection time hence getting very
large, as visible in the relative time-space chart, at higher voltages
it just slightly expands and then shrinks back. Moreover, the time of
maximum expansion is reached earlier at higher voltage
(Fig.~\ref{fig:2}F). It should also be noted that this happens
regardless of the fact that the electron cloud starts out with an
initial width $\sigma(t=0)$ that is larger for higher voltage. This
latter is just a featuring detail, which can be considered a real effect
due to the interplay of pump pulse width with a larger speed, partially
augmented by a blurry artefact due to the probe width (see Methods).
The cloud-size dynamics are worth their deepening, even here in a
linear (i.e., not repulsive) regime. The Gaussian shape of the cloud is
in agreement with basic models which include both thermal diffusion and
Coulombic repulsion
\cite{gatti_dynamics_1987,benoit_simulation_2009} to describe the
charge dynamics in semiconductor detectors. Such models provide a
Gaussian peak with a moving centre and an increasing width proportional
to diffusion and repulsion, under a uniform electric field. In the
present case of low-power experiments, repulsion should be ruled out,
and the only mechanism for the cloud size's growing should be carrier
diffusion, which predicts a size increase following a square root with
time $\sigma = (2Dt)^{1/2}$ (with $D$ the diffusion coefficient). In
any case, no dynamic reduction of the size is predicted. We have hence
to observe that the observed non-uniformity of the electric field is
effective. In a decreasing field, the front tends to decelerate earlier than the tail. Hence, the apparent quasi-parabolic evolution
of $\sigma(t)$ comes from the competition between the diffusion and the
decreasing electric field.

CDPI provides full access to basic transport properties without the
need for spatial averaging and the approximations implicit in any other
experimental evaluation. As a harsh-sounding example, the carrier
mobility is typically evaluated using time-of-flight measurements and
the following relation: $\mu_{\mathrm{ave}} = L^2 / t_c V$. The
collection time $t_c$ is extracted from the decaying currents with a
certain degree of arbitrariness, while a uniform electric field $V/L$
is assumed. CDPI provides trajectories and thus an accurate estimation
of local velocity at the local electric field. However, from the
non-approximated expression
$\mu = L / \int_{0}^{t_c} E(y_0, t)\,\mathrm{d}t$, we have extracted
values averaged over all different voltages around
$\mu = (1076 \pm 73)$ cm$^{2}$ V$^{-1}$ s$^{-1}$, very similar to those
calculated assuming a constant electric field, i.e.
$\mu = (1074 \pm 43)$ cm$^{2}$ V$^{-1}$ s$^{-1}$, these values being in
good agreement with the literature \cite{suzuki_effect_2018}.

\section*{Electron clouds at large optical irradiation}
We now study the field and charge carriers' dynamics observed under
larger levels of optical irradiation. Here the pump beam power density
$P$ varied along more than three orders of magnitude ($P_1$--$P_8$
going from approximately 1 $\mu$W cm$^{-2}$ to 3900 $\mu$W cm$^{-2}$),
while the applied bias voltage was always $V = 200$ V.

\begin{figure*}[htbp]
  \includegraphics[width=0.9\textwidth]{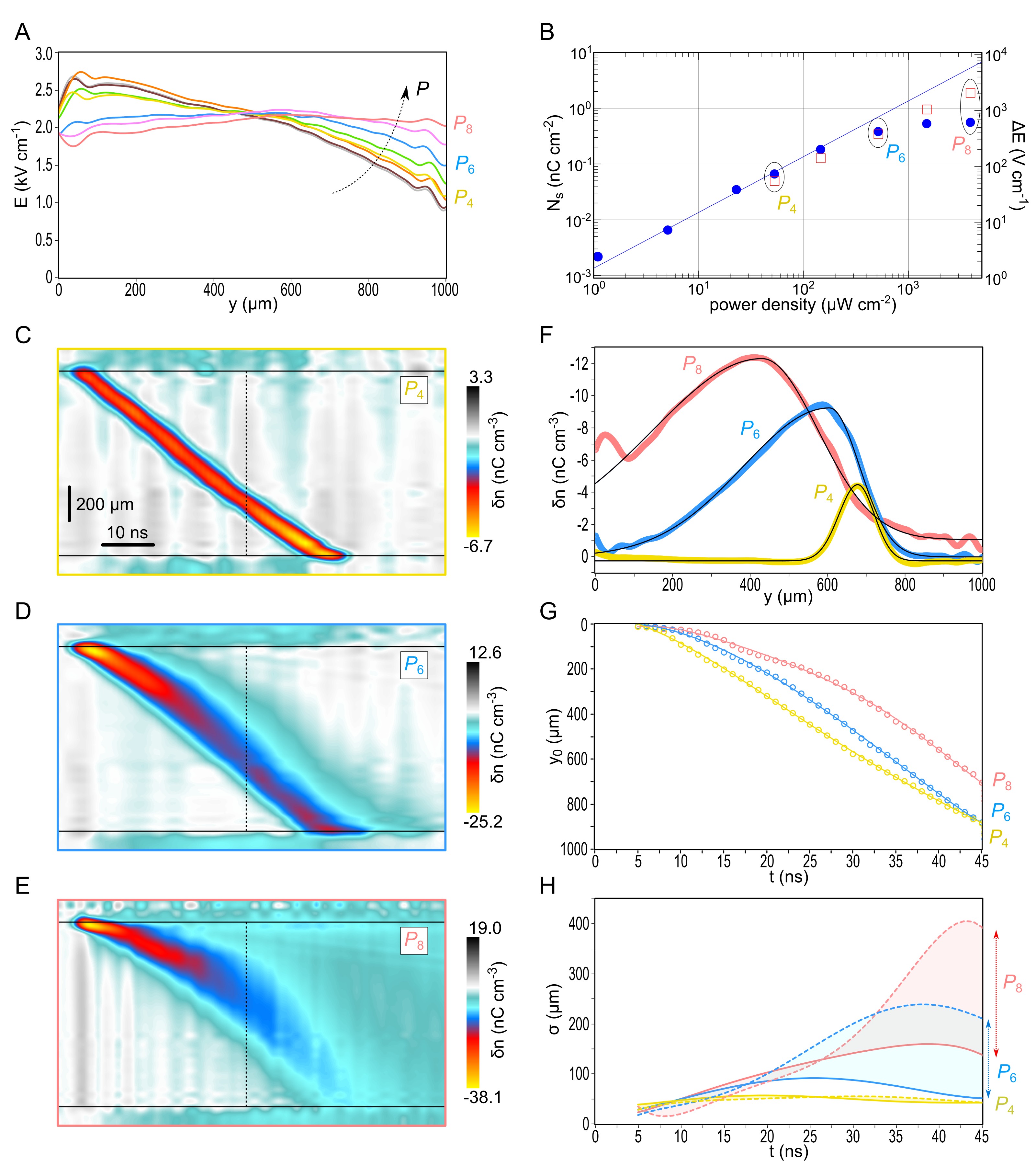}
  \caption{\label{fig:3}%
    \textbf{CDPI experiments at increasing optical power.}\\
    (A)~Electric field profiles at initial time for different optical
    pump power densities (from $P_1=1$ $\mu$W cm$^{-2}$ to
    $P_8=3900$ $\mu$W cm$^{-2}$, over the 1 cm$^{2}$ area of the
    device).
    (B)~Estimated travelling charge (filled blue circles) for the
    eight values of increasing power. The charge is evaluated in terms
    of the electric-field difference $\Delta E$ at the two sides of the
    clouds (at intermediate time). The collected charge is retrieved
    instead by time integrating the approximated current (empty red
    squares).
    (C,D,E)~Time-space charts of the carriers' cloud
    $\delta n_{\mathrm{displ}}(y)$ at three optical powers ($P_4$, $P_6$ and
    $P_8$, respectively). They are retrieved by applying Eq.~(3) to
    the electric field charts (please note that here the carrier cloud
    is reported in terms of charge density, in nC cm$^{-3}$). The time
    range is 72 ns, and the time step is 1 ns. The inner top and
    bottom horizontal solid lines represent the cathode and anode
    positions, respectively.
    Spatial and temporal scale bars 
    represent 200 $\mu$m and 10 ns, respectively.
    (F)~Spatial profiles of $\delta n_{\mathrm{displ}}(y)$ for the same optical
    powers (yellow, blue and red for $P_4$, $P_6$ and $P_8$,
    respectively) taken at $t=35$ ns, corresponding to the vertical
    dashed line in the panels (C,D,E). The bi-Gaussian fits are
    superimposed (solid black lines).
    (G)~Peak position $y_0$ extracted from the bi-Gaussian fits during
    time (the data are empty dots; solid lines are a guide for the
    eye).
    (H)~The extracted front and tail widths (solid and dashed lines,
    respectively) as a function of time.%
  }
\end{figure*}

First, it is to be noted that a prolonged exposition to optical
irradiation has been used (around 20 min) to achieve a steady state
(Fig.~S2) before starting the Pockels acquisitions. This was necessary
because, under high optical irradiation, the electric field
perturbation remains much longer than the transit time, and even longer
than the time interval associated with the repetition rate of 25 kHz.
This effect is analogous to the flux-dependent polarization, critical
in CdTe detectors under high X-Ray fluxes and it basically depends on
the modification of deep trap occupation \cite{bale_nature_2008}. The
modification is visible in the $E$-field profiles at $t = 0$ shown in
Fig.~\ref{fig:3}A, where the electric field undergoes a progressive
bending. In particular, the field reduces under the irradiated cathode,
while it increases close to the anode, upon increasing the incident
optical power. Hence the meaning of these profiles is that they
represent the initial electric field that will both guide and be
perturbed by the moving electron cloud.

A significant quantity to the extent of evaluating the induced dynamics
and operational features is the collected charge, which can be simply
obtained from the electric field step associated to the localized
drifting electron cloud, according to
$\Delta E = \frac{q}{\varepsilon} N_s$, where $N_s$ is the areal charge
density. The value $\Delta E$ (taken at $t = 35$ ns into the dynamics)
is reported in Fig.~\ref{fig:3}B as a function of the incident optical
power, highlighting a reasonable linear relation under low to moderate
optical powers. However, it saturates to a value around 500 V cm$^{-1}$
at the larger powers. This value is substantially lower than the value
of the electric field at the cathode, of the order of
2000 V cm$^{-1}$ (see Fig.~\ref{fig:3}A), which means that even under
high optical power the electric field under the irradiated electrode
does not drop to zero at the time the electron cloud is
photogenerated. The relation between carrier dynamics, electric field,
trapping, contact properties and how these all affect the currents, has
been deeply investigated in the past
\cite{weisz_electrode-limited_1968,batra_photocurrents_1970,many_theory_1962,gibbons_transient_1968,christen_effect_2016},
especially in the frame of strong carrier injection. For ohmic
contacts, it is predicted that the space-charge-limited (SCL) regime
can be eventually achieved, i.e., the electric field is nulled at the
electrode where strong photogeneration or high carrier injection
occurs. The limit case of full absence of electron injection, i.e., a
blocking contact, results in a limited electric field reduction near
the blocking contact where carriers are optically generated. This case
is referred \cite{papadakis_theory_1967} to as space-charge perturbed
(SCP). We argue that the limited electric field reduction that we
observe under optical pulsed excitation can be attributed to the upward
energy band bending at the Pt/CdTe interface, which makes such contact
slightly hole injecting \cite{farella_study_2009} thus impeding full
electron injection from the cathode and the nulling of electric field
nearby. On the side of low optical power in Fig.~\ref{fig:3}B, it
is quite remarkable that we can appreciate electrical field changes
down to 2 V cm$^{-1}$ over a baseline of the order of few thousand
V cm$^{-1}$. This corresponds to detecting electron clouds with sheet
density down to $7\times 10^{6}$ cm$^{-2}$. Such a concentration
translates, for a typical cloud width of 40 $\mu$m, to the extremely
low volumetric density of $3.5\times 10^{9}$ cm$^{-3}$ of
photogenerated carriers, just above the intrinsic carrier
concentration.

The time-space dynamics of the carrier appears in the charts in
Fig.~\ref{fig:3}C,D,E for three different values of optical power
density (respectively, $P_4 = 57$ $\mu$W cm$^{-2}$,
$P_6 = 520$ $\mu$W cm$^{-2}$ and $P_8 = 3900$ $\mu$W cm$^{-2}$). These
panels evidence both a distortion of the trajectory (linear in the low
power case) and a strong enlargement of the cloud size, the last one
predicted in the literature by numerical simulations
\cite{suzuki_effect_2018,castoldi_simulation_2013}. The wider size of
the cloud also implies that the associated $J_{\mathrm{dis}}$ is not
conformally mapped to the charge cloud, for this reason, we recur here
to the $\delta n_{\mathrm{displ}}(y)$ visualization. It can be
appreciated from the charts that the moving cloud starts out at a lower
speed with increasing optical power, despite accelerating at later
times. This trend is in qualitative agreement with the electric field
profile across the device (Fig.~\ref{fig:3}A). A better look at its
position and shape is offered by the vertical crosscuts of the charts
at fixed time ($t = 35$ ns of the dynamics), shown in
Fig.~\ref{fig:3}F. The peak has evidently slowed down, and its spatial
distribution enlarges with increasing optical power (Video S4).
Furthermore, at low optical power (up to $P_4$) the spatial
distribution is symmetric and, as we have already seen, well described
by a Gaussian function (superimposed fit, solid black line). On the
contrary, under high optical powers ($P_6$ and $P_8$) a bi-Gaussian
function is needed to follow the data, bearing different widths for the
front ($\sigma_f$) and for the back ($\sigma_b$) of the cloud. The
bi-Gaussian fits carried out during the drift of the electron cloud
allow us to visualize the trajectory change by looking at the position
of the peak $y_0$ in time, as reported in Fig.~\ref{fig:3}G. The
trajectory starts to change at powers larger than $P_4$, with an
appreciable slowing down during the initial instants. It should be
warned though that the large extension and deformation of the electron
cloud limit the meaning of trajectory. Concerning the spatial shape of
the cloud, we report in Fig.~\ref{fig:3}H the evolutions of
$\sigma_f$ and $\sigma_b$ in time (solid and dashed lines,
respectively). A systematic trend can be identified. At low optical
power the two widths basically overlap during
motion, indicating a Gaussian shape and negligible asymmetry. Their
value grows just slightly before decreasing (as also seen in
Fig.~\ref{fig:2}F). At larger powers the cloud gets wider and growing
till longer times, with a maximum frontal width reaching 80 $\mu$m (at
20 ns) and 160 $\mu$m (at 32 ns), for $P_6$ and $P_8$, respectively.
More strikingly, the tail width gets even larger and at increasingly
later time, up to 240 $\mu$m (at 33 ns) and 400 $\mu$m (at 38 ns), for
$P_6$ and $P_8$, respectively. In summary, it is evident that increasing
the optical excitation power causes the cloud of electrons to spread
out over a large part of the sample in agreement with previous
simulations \cite{suzuki_effect_2018,castoldi_simulation_2013},
becoming increasingly asymmetric with a tail that is larger than the
front.

When increasing the incident optical power, Coulombic repulsion plays a
dominant role. Compared to thermal diffusion, Coulombic repulsion
predicts again a Gaussian distribution in 3D configurations but with a
width proportional to $(Nt)^{1/3}$, where $N$ is the total number of
photogenerated carriers per pulse. Calculations
\cite{gatti_dynamics_1987,benoit_simulation_2009} are for a
spherical/ellipsoidal carrier cloud shape, while in our case the
carrier cloud is a 2D sheet moving in one direction. This is still
expected to expand with a $(Nt)^{1/3}$ dependence in such a direction,
but with a higher multiplicative factor to account for enhanced
repulsion due to parallel lines of force. The cloud broadening effect
of repulsion is clear from Fig.~\ref{fig:3}H, where both $\sigma_f$ and
$\sigma_b$ exhibit a net increase with time and the incident optical
power. However, diffusion and Coulombic repulsion alone cannot account
for the pronounced asymmetry of the cloud. Trapping/detrapping
processes provide an additional mechanism capable of modifying both the
carrier concentration and the spatial evolution of the cloud. In
particular, the tail observed in the electron distribution, together
with the delayed charge evolution discussed below, suggests that
trapping becomes relevant under high optical excitation. Generally, at
operative electric fields ($> 600$ V cm$^{-1}$), electron trapping is
considered negligible because the associated lifetime is of the order
of 1 $\mu$s, much longer than the carrier transit time. Nevertheless,
nonlinear trapping-related effects under intense optical excitation
have been reported in CdTe and related compounds
\cite{suzuki_effect_2018,wurm_study_1994,kamieniecki_effect_2014}.
Additional insights are provided by Fig.~\ref{fig:3}B. As the optical
power increases, the charge $N_s$ evaluated from the moving electron
cloud during drift (blue points) becomes progressively smaller than
that obtained by integrating the total current transient over a longer
time window (Eq.~\eqref{eq:4}, red open squares). This difference
indicates that an increasing fraction of the photogenerated charge is
temporarily removed from the mobile electron population during
transport and is recovered only at longer timescales through delayed
release processes. Such behaviour is consistent with trapping followed
by detrapping, rather than with a permanent loss of charge. Additional
experiments performed at variable temperature (not shown here) indicate
that the long-time tail of the electron-cloud evolution is thermally
activated, providing further evidence for a detrapping process.

A peculiar feature is observed in Fig.~\ref{fig:3}E (curve $P_8$),
where a secondary peak appears very close to the cathode ($y = 0$).
This peak persists for times longer than the carrier collection time
and can be attributed to a localized fixed charge contribution
$n_t(y,t)$ associated with traps near the optically irradiated cathode.
In this regard, fabrication processes of CdTe detectors are known to
introduce a defective region, approximately 2 $\mu$m thick, below the
contact, commonly referred to as a ``dead layer'', which can prevent
full charge collection \cite{zahraman_study_2006}. The influence of
both distributed and localized defects on the observed charge profiles
will be further investigated through numerical simulations. The
possibility offered by CDPI to distinguish the contributions of mobile
carriers and trapped charge represents an additional advantage of the
technique. To the best of our knowledge, this is the first direct
visualization of the spatiotemporal evolution of a photogenerated
carrier cloud inside an operating electronic device, including the
emergence of its high-excitation asymmetry.

\section*{Current transients}
\begin{figure}[thbp]
  \includegraphics[width=0.9\columnwidth]{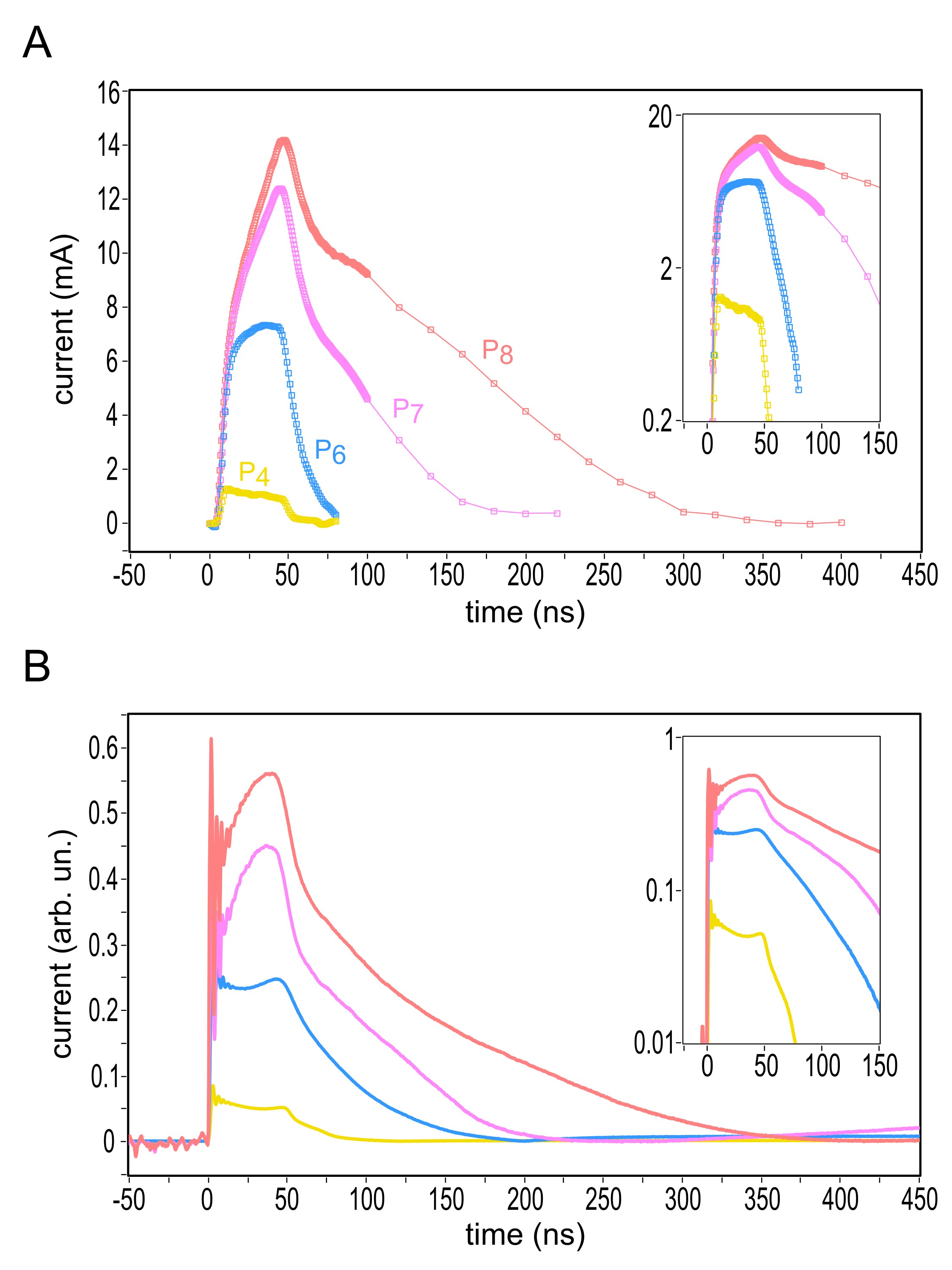}
  \caption{\label{fig:4}%
    \textbf{Current dynamics.}\\
    (A)~Experimental current versus time at different powers
    calculated from the CDPI experiments using Eq.~(4).
    (B)~Independent current transients (CT) measurements by means of
    an external oscilloscope. The insets in the two panels show the
    currents on a semi-log scale.%
  }
\end{figure}

Albeit approximate, Eq.~\eqref{eq:4} provides a straightforward and
compact expression for the current transient, avoiding the use of noisy
electric field derivative of Eq.~\eqref{eq:2}, but only relying on their
values at the electrodes. Figure~\ref{fig:4} shows a comparison of the
current transients for different optical powers. These are either
calculated from the CDPI experiments by using Eq.~\eqref{eq:4}
(Fig.~\ref{fig:4}A) or taken from independent CT (current transient)
experiments (Fig.~\ref{fig:4}B) measured by means of an oscilloscope,
see also Methods. While the transient edges generally align well (at
around 50 ns, corresponding to the transit time at 200 V), the peak
occurs slightly earlier in the CT experiments conducted at the highest
optical powers. The SCP \cite{weisz_electrode-limited_1968} and SCL
\cite{many_theory_1962} models predict a peak occurring at
approximately 0.8 of transit time, which is consistent with the
observations made in the CT experiments. The slope of the transient
(the part before the edge at 50 ns) changes in both kinds of experiments
from negative to positive when increasing the optical power. These time
slopes are a footprint of either decelerating or accelerating carrier
clouds, consistent with the shape of the spatial profiles in the
background electric field as seen in Fig.~\ref{fig:3}A. After the edge,
a decay happens in both experiments, that is longer with increasing
optical power. The development of such a temporal tail in the current
mirrors the delayed component already observed in the carrier-cloud
evolution, providing an independent consistency check of the CDPI
interpretation. Indeed, trapping/detrapping events can delay the arrival
of carriers at the collecting electrode far beyond the regular transit
time.

At the start of the transients and for few ns, CT experiments show a
fast initial peak (even though affected by some instrumental signal
ringing). Such a component is missing in the CDPI transients and is
induced, as will be confirmed by numerical simulations, by the
photogenerated holes. Since holes are photogenerated near the cathode
and rapidly collected, they provide a fast initial current contribution
but also affect the electric field at the cathode with a sign opposite
to that of electrons. The apparent initial current deficit observed in
the CDPI transients therefore originates from applying the
single-carrier approximation of Eq.~\eqref{eq:4} to a situation where
both carrier types contribute to the electric-field evolution.

In summary, within its limits, the approximate current of
Eq.~\eqref{eq:4} validates the central result of CDPI by establishing a
direct correspondence between the observed internal electric-field
dynamics and the independently measured external current. Under
constant applied bias, the spatially integrated displacement-current
contribution vanishes identically. Consequently, the external current
transient reflects the spatially averaged conduction current, whereas
CDPI enables reconstruction of the spatial distributions of both
conduction and displacement current contributions within the device.
Remarkably, this all-optical approach simultaneously provides access to
the evolving electric field, the corresponding charge distribution, and
the local current components, while correctly predicting the externally
measured current transient. In this respect, CDPI provides an
experimental counterpart to the physical picture underlying the
generalized Ramo-Shockley theorem \cite{cavalleri_extension_1971}.
While the generalized Ramo-Shockley formalism relates the externally
induced current to charge motion through the weighting-field concept,
CDPI directly measures the evolution of the internal electric field,
from which the corresponding charge distribution and local current
components are reconstructed.

\section*{Numerical simulations}
To improve our understanding of experimental phenomenology, we performed
numerical simulations that captured most of its features and results.
The time-dependent drift-diffusion equations were solved by starting
from an established model \cite{cola_optical_2022,cola_relation_2016}
for the bulk of the CdTe and the contacts, here updated for transient
conditions (see Methods and Table~S2 for details).

\begin{figure*}[t]
  \includegraphics[width=0.9\textwidth]{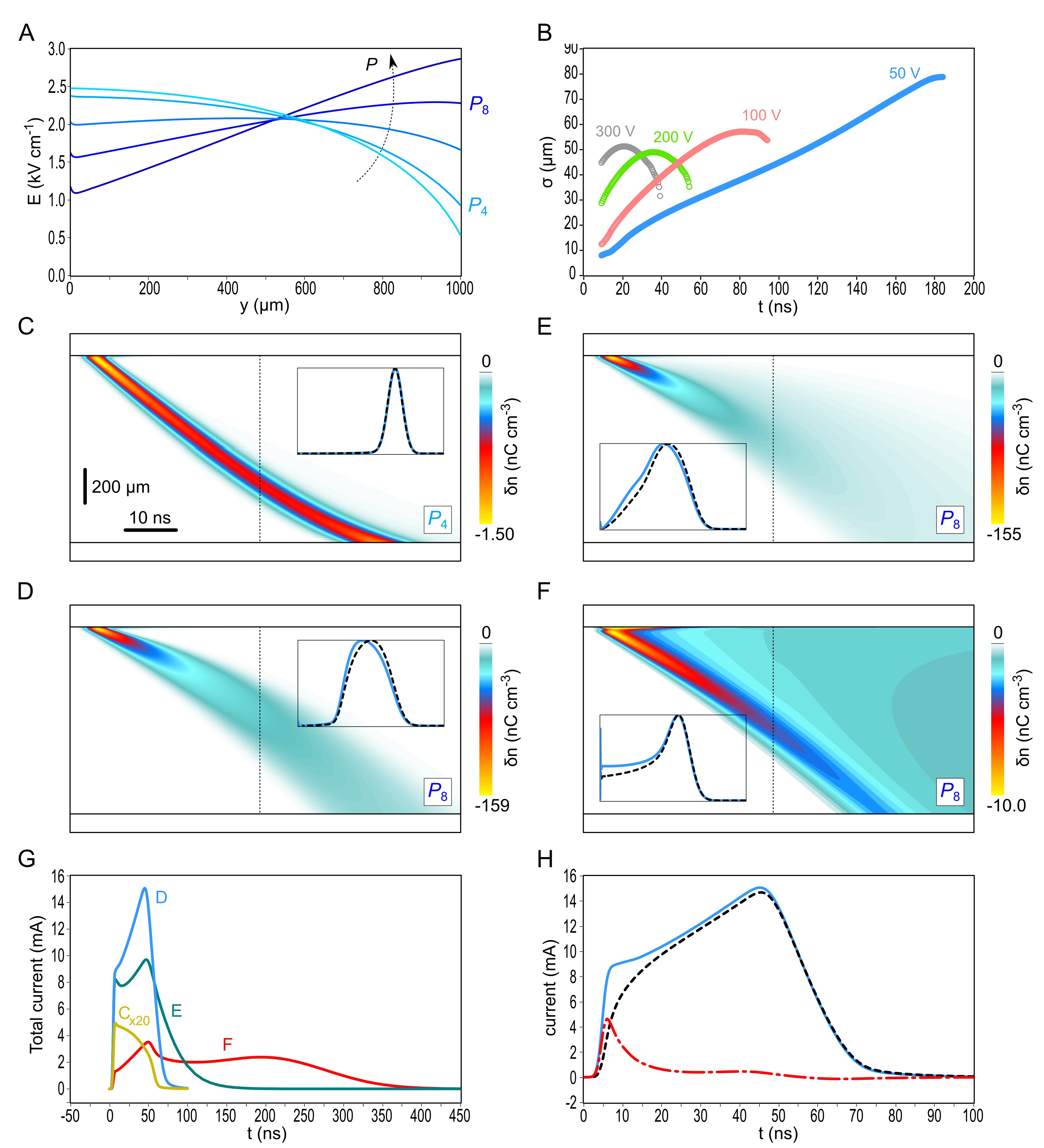}
  \caption{\label{fig:5}%
    \textbf{Numerical simulations.}\\
    (A)~Initial field profiles prepared at different optical pump
    powers. The two labelled profiles correspond to the $P_4$ and
    $P_8$ power densities of the experiments.
    (B)~Evolution of the cloud size extracted from the Gaussian fit of
    the simulations at different bias voltages and low-power optical
    irradiation.
    (C,D)~Time-space charts of the carrier cloud
    $\delta n_{\mathrm{displ}}(y,t)$ in a low-power case (C) and high-power
    case (D), under an applied voltage of 200 V (please note that here
    the carrier cloud is reported in terms of charge density, in
    nC cm$^{-3}$).
    Spatial and temporal scale bars 
    represent 200 $\mu$m and 10 ns, respectively.
    (E,F)~Time-space charts of the carrier cloud using the same high
    optical pump power of panel (D) but adding a trapping level. The
    two cases refer to a shallow level homogeneous in the bulk depth
    (E) and a surface defect level localized in the first few $\mu$m
    from the cathode (F). The insets in panels (C,D,E,F) show the
    normalized profiles at fixed time ($t=35$ ns into the dynamics,
    dashed vertical line in the charts), highlighting the
    correspondence in shape of the charge [$\delta n_{\mathrm{displ}}(y)$,
    solid blue lines] and the displacement current itself
    [$J_{\mathrm{dis}}(y)$, dashed black lines]. The spatial and time ranges
    of the dynamics are 1000 $\mu$m and 72 ns, respectively, in all
    the four charts.
    (G)~Total current transients $J_{\mathrm{tot}}(t)$ for simulations of
    panels (C,D,E,F).
    (H)~Comparison of $J_{\mathrm{tot}}(t)$ (solid blue line) and
    $J_{\mathrm{approx}}(t)$ (dashed black line) in the same high-power case as
    panel (D). The panel also shows their difference (dash-dot red
    line), evidencing the initial hole contribution.%
  }
\end{figure*}

In order to mimic the experiments (see protocol in Fig.~S2),
preliminary and prolonged (15--20 min) optical expositions to prepare
the system are simulated, assuming an effective, time averaged
intensity for the optical pump. After such conditioning, a balance
between the long trapping and de-trapping times of the deep level
occurs which can modify the space charge distribution, and thus the
electric field profile with respect to dark. Like in Fig.~\ref{fig:3}A
for the experiments, simulations in Fig.~\ref{fig:5}A show that the
profiles are not affected in case of low irradiation while, for high
pulse levels, the electric field is increasingly damped on the cathode
side. Also in simulations, such profiles represent the initial electric
field before the dynamics.

The simulations confirm that, under low optical irradiation, the
electron cloud retains a well-defined Gaussian spatial distribution at
any instant and for the entire drift of the cloud, when different
applied voltages are applied. We have performed simulations
corresponding to the experiments of Fig.~\ref{fig:2} and subsequently
applied Gaussian fits on the retrieved $\delta n_{\mathrm{displ}}(y,t)$
clouds during motion. The widths $\sigma(t)$ as a function of time are
reported in Fig.~\ref{fig:5}B for different voltages, showing a good
agreement with the corresponding CDPI results of Fig.~\ref{fig:2}F.
While at the lowest voltage the cloud size increases along all the
transit, for larger voltages $V > 50$ V it shows a maximum $\sigma$ in
the order of 50--60 $\mu$m before getting reduced. The maximum width is
reached earlier for larger cloud speeds, i.e., for higher applied
voltages. This confirms the interplay between electron diffusion and a
decreasing electric field, where the latter tends to squeeze the
electron cloud. It is important to note that also in the case of
numerical simulations, the optical pump width (1 ns) working together
with the cloud speed plays a role in the charge cloud broadening. This
is evident from the voltage-dependent size of the cloud at the
beginning of the dynamics shown in Fig.~\ref{fig:5}B, which can also be
seen in the CDPI result (Fig.~\ref{fig:2}F).

In Figs.~\ref{fig:5}C and \ref{fig:5}D, we report the time-space
charts of the charge cloud corresponding to low and high optical
irradiation, respectively. These results should be compared with the
CDPI measurements shown in Fig.~\ref{fig:3}C (low excitation) and
Fig.~\ref{fig:3}E (high excitation). Similarly to the experiments, the
two optical regimes exhibit opposite trajectory bending, with
deceleration at low excitation and acceleration at high excitation. The
latter case also shows a pronounced cloud broadening. While the change
in carrier motion follows the inversion of the electric-field profiles
shown in Fig.~\ref{fig:5}A, the cloud expansion originates from Coulomb
repulsion, which is intrinsically included in the model.

The cross-sectional profiles of the $\delta n_{\mathrm{displ}}(y,t)$ distributions 
extracted from the time-space maps of Figures \ref{fig:5}C and \ref{fig:5}D 
are reported in the
corresponding insets, allowing the evolution of the cloud shape to be
evaluated at selected times. At low optical power (Fig.~\ref{fig:5}C),
the simulations reproduce well the experimentally observed Gaussian
carrier cloud, indicating that the basic transport model adequately
describes the carrier dynamics under these conditions. At high optical
power (Fig.~\ref{fig:5}D), however, significant discrepancies emerge.
Strikingly, the simulated asymmetry is opposite to that observed
experimentally (see $P_6$ and $P_8$ profiles in Fig.~\ref{fig:3}F),
with the front side extending slightly more than the rear side.

The experimental asymmetry is therefore a distinctive feature that
cannot be reproduced by the basic model, which excludes appreciable
trapping. The model already accounts for the electric-field
redistribution produced by space charge, Coulomb repulsion, and the
finite optical absorption profile. None of these effects reproduces the
experimentally observed high-excitation asymmetry. Only after
introducing trapping/detrapping does the simulated carrier-cloud
evolution become consistent with the experimental observations.

To identify the additional mechanism required by the experiments, we
introduced an additional shallow electron trapping level in the
simulations performed under high optical excitation. Two physically
motivated trapping scenarios were considered: a spatially uniform bulk
distribution and a localized near-electrode distribution
(Figs.~\ref{fig:5}E and \ref{fig:5}F), both under the same excitation conditions as
Fig.~\ref{fig:5}D. In both cases, the cloud broadens further, mainly
due to an extension of the rear part, reproducing the asymmetry
observed by CDPI. The delayed tail is visible in the profiles shown in
the insets and exhibits an approximately bi-Gaussian shape for the
uniform trap distribution (Fig.~\ref{fig:5}E). An even longer spatial
tail develops for the localized trap case (Fig.~\ref{fig:5}F),
characterized by an almost flat rear plateau.

The inclusion of a shallow trapping level simultaneously reproduces the
delayed spatial tail of the carrier cloud and the corresponding
temporal tail of the current transient observed experimentally. The
latter naturally follows from the delayed arrival of trapped carriers
released after the transit of the main electron packet.
Figure~\ref{fig:5}G reports the simulated total current for the cases
shown in Figs.~\ref{fig:5}C--\ref{fig:5}F. Without trapping 
(cases shown in Figs.~\ref{fig:5}C and \ref{fig:5}D), 
the current transient remains relatively sharp ($< 70$ ns). Conversely, a
delayed component appears for the shallow bulk trap 
(case shown in Fig.~\ref{fig:5}E), while the
localized trap (case shown in Fig.~\ref{fig:5}F) 
produces slower dynamics, with the main peak
followed by a secondary feature extending up to 350 ns.

The experimental phenomenology may therefore involve a combination of
distributed and localized trapping effects, which could be further
refined in future modelling. Rather than identifying a unique
microscopic model, these simulations demonstrate how CDPI provides
direct experimental access to transport mechanisms that have previously
been inferred mainly from externally measured current transients, which
integrate all spatial contributions into a single electrical
observable.

We want to point out that numerical modelling allows us to validate a
fundamental statement in the CDPI technique presented before. The
possibility of evaluating the electron cloud
$\delta n(y,t)$ from the displacement-current
contribution, through the equivalent quantity
$\delta n_{\mathrm{displ}}(y,t)$ introduced in Eq.~\eqref{eq:3}, has
been systematically assessed under all the conditions discussed above.
Furthermore, we also show in the inset of
Figs.~\ref{fig:5}C--\ref{fig:5}F the difference in the moving charge
[$\delta n_{\mathrm{displ}}(y)$, solid lines] and displacement current
itself [$J_{\mathrm{dis}}(y)$, dashed lines]. While only a slight
deviation is noted for high-power cases, the perfect overlap in the low
power case confirms that the displacement-current contribution provides
a reliable reconstruction of the charge-cloud shape and dynamics under
localized-cloud conditions.

Finally, we analyze the role of the photogenerated holes, whose
presence was inferred from CT experiments in Fig.~\ref{fig:4}B. In
terms of conduction currents and their spatial distribution, simulations
highlight that holes become relevant, and even dominant when very close
to the cathode (Fig.~S3). Looking at the current transients, such holes
result in an additional contribution at the very beginning of the
dynamics. They can eventually manifest in a peak, which can be barely
perceived in Fig.~\ref{fig:5}G. Their contribution becomes more evident
when the time range is expanded and comparing the retrieved
$J_{\mathrm{tot}}(t)$ and $J_{\mathrm{approx}}(t)$ transients in
Fig.~\ref{fig:5}H (solid blue and dashed black, respectively, for the case
of Fig.~\ref{fig:5}D). In fact, the $J_{\mathrm{approx}}(t)$ expression has been
obtained from simulations by assuming no photogenerated holes. The
difference between $J_{\mathrm{tot}}(t)$ and $J_{\mathrm{approx}}(t)$
(dash-dot red curve) is showing an appreciable initial peak that is
precisely due to the holes' contribution. The difference also points
out an extended but very weak later bump. This is due to other
approximations involved when deriving Eq.~\eqref{eq:4}.

More generally, these results highlight another important capability of
CDPI. In principle, electrons and holes can be separately retrieved
once their spatial distributions become sufficiently separated, since
their opposite charges generate electric-field perturbations that
evolve in opposite directions under the applied bias. This capability
is only temporarily limited when the two carrier populations overlap
spatially, as immediately after photogeneration or during crossing
events following injection from opposite electrodes, because their
contributions to the electric field partially compensate each other
according to Gauss's law. The possibility of directly resolving the
spatiotemporal evolution of oppositely charged carrier populations
represents one of the most distinctive capabilities of CDPI for
investigating charge transport in semiconductor devices.

\section*{Conclusions}
We have implemented CDPI, a sensitive, non-invasive and time-resolved
pump-probe imaging approach for electro-optic materials and
optoelectronic devices exhibiting a measurable Pockels response. CDPI,
here applied to CdTe radiation detectors, is based on the fundamental
principle that any carrier excess dynamically affects the internal
electric field. As such, the method enables direct and quantitative
visualization of optically generated electron clouds and reconstruction
of the associated local conduction and displacement-current
distributions. Since Pockels effect is related to the medium
polarizability, the modulation is extremely fast and thus CDPI can in
principle access carrier dynamics on timescales determined by the
optical excitation and detection scheme rather than by the intrinsic
response time of the Pockels effect. Moreover, since the effect is
probed by a not-absorbed optical beam, CDPI is not invasive.

Besides the applicative interest, fundamental carrier transport
properties, like drift, diffusion, repulsion, trapping/detrapping, and
electrode injection can be accessed directly from spatially resolved
measurements of the electric-field evolution, reducing the need for
indirect inference from spatially averaged electrical measurements.
The present results refer to electrons, but CDPI is also suitable to
investigate hole clouds, depending on which of the two electrodes
is irradiated. We have found that under different applied voltages and
low optical intensity, the Gaussian shape of the electron cloud is well
maintained during drift, with the width first increasing due to diffusion
and then shrinking due to decreasing electric field profiles. On the
front of sensitivity, we have detected electric field differences of
about 2 V cm$^{-1}$ associated to the drift of low-concentration
electron clouds, just above the intrinsic level.

Conversely, high optical excitation intensities allow us to investigate
perturbed space charge regimes, which have traditionally been
investigated mainly through externally measured current transients.
Notably, in addition to significant expansion of the electron cloud due
to Coulomb repulsion, we have observed pronounced tails in the moving
clouds due to detrapping processes, as consistent with independent
current measurements. We have deduced that trapping in CdTe detectors
is enhanced by optical intensity and mostly associated with a defective
layer possibly close to the surface. This leads us to another crucial
point that we have independently verified: our all-optical approach
correlates the dynamics of internal charges with the current flowing in
an external circuit. This correlation between charge dynamics and
external current response, achieved through electro-optic field
mapping, provides a new framework for studying charge transport in
operating devices.

A standard drift-diffusion simulator, equipped with a well-established
model, can capture all the observed CDPI features. Besides validating
our approach, this substantially improves control of present CdTe
devices from a microscopic point of view to the level of device
performance. Furthermore, given that CDPI can easily be extended to two
dimensions, it paves the way for novel or improved designs in devices
where charge collection is critical, such as detector matrices.

Finally, CDPI shows potentialities along several directions.
\textbf{(1)~Materials.} Beyond CdTe, CDPI can be extended to other
noncentrosymmetric semiconductors, perovskites, nonlinear optical
organic materials and established electro-optic platforms, provided
that a measurable Pockels effect is available.
\textbf{(2)~Time resolution.} This can be further advanced, being here mainly
limited by pump-probe laser parameters such as pulse energy, pulse
duration, and timing jitter.
\textbf{(3)~Spatial dimensionality and resolution.} CDPI can be adapted to probe
microscopic devices, operated in reflection as well as transmission,
and extended from the current implementation to fully two-dimensional
distributions.
\textbf{(4)~Devices.} The technique is broadly suited to investigating devices
based on electro-optic materials in which is involved the propagation
of either electrically or optically generated excess carriers.

\section*{Methods}

\subsection*{Experimental setup}
The radiation detectors (Acrorad Co., Ltd., Okinawa, Japan) used in the
experiments are based on CdTe:Cl semi-insulating crystals (resistivity
$\rho = 6$--$7\times 10^{9}$ Ohm cm) \cite{cola_relation_2016}, slightly
p-type of size $10~\times~10~\times~1~\text{mm}^{3}$. The nominal drift
electron and hole mobilities are
$\mu_e = 1050$ cm$^{2}$ V$^{-1}$ s$^{-1}$ and
$\mu_h = 80$ cm$^{2}$ V$^{-1}$ s$^{-1}$ respectively. The Pt contacts,
partially transparent, cover the whole upper and lower square surfaces,
parallel to the (111) crystal planes. Bias voltage is applied by a
B2987B Keysight Electrometer. Four nominally equal detectors were used
in this work, and the results were highly comparable. The sample is
placed on a thermal stage whose position is carefully controlled along
six degrees of freedom by multiple translational and angular stages.
Measurements have been carried out in air at room temperature.

Two optical beams are impinging on the sample, in a pump and probe
configuration, coming from two independent pulsed lasers both of
measured width around 1.0 ns. The pump beam impinges vertically on the
top semi-transparent contact, while the probe beam is perpendicular to
the pump and it impinges on the lateral side of the sample. The probe
pulse is electrically triggered by the pump pulse through a pulsed delay
generator (Highland Technology T564). The jitter between the two
optical pulses is less than 100 ps, thus resulting in an overall
temporal resolution mainly limited by the probe pulse width (1 ns).

The pump beam is produced by a supercontinuum white laser (Leukos,
Samba), repetition rate 25 kHz, which is spectrally selected by a
band-pass filter (780--830 nm) to be immediately above the energy
bandgap ($E_g = 1.43$ eV, corresponding to $\lambda_g = 840$ nm) for a
penetration length short ($\approx 1$ $\mu$m) with respect to the
sample thickness of 1 mm (Fig.~S1). The resulting optical power density
is varied by means of neutral filters in the
1--3900 $\mu$W cm$^{-2}$ range. The wavelength of the laser probe beam
(customized by Eikonal Optics) is 980 nm, well below the energy bandgap
and is thus not absorbed by the sample. Its power is also small enough
(tens of $\mu$W) not to induce additional electro-optic effects. This
allows us to probe the internal electric field by means of the Pockels
effect in a transmission configuration. After a half-wave plate, the
probe pulse is linearly polarized at 45$^{\circ}$ with respect to the
vertical $y$-direction and experiences the electric field-induced
birefringence when crossing the sample. The effect is detected due to a
second polarizer (analyzer) oriented at $-45^{\circ}$, placed behind
the device (cross-polarizer configuration). A half-wave plate is
placed before the first polarizer to maximize the optical intensity at
disposal. The images $P(x,y)$ of the transmitted intensity are recorded
by a cooled high-sensitivity monochrome camera (Q Imaging, EXi Blue)
equipped with a 1$\times$--6$\times$ zoom lens (Edmund Optics,
VZM\texttrademark{} 600i Zoom Imaging Lens). The system is suitable to
image the 1 mm detector thickness onto a 1392 $\times$ 1040-pixel area.
The camera pixel size is 6.45 $\mu$m, and the integration time ranges
from 3 s to 5 s. The spatial resolution of the setup in the used
experimental configuration is limited by different factors: the
numerical aperture NA of the optical zoom lens at the employed
magnification (5$\times$ magnification, object-space NA $\approx 0.05$,
corresponding to a diffraction-limited resolution of
$\approx 6.7$ $\mu$m), the horizontal optical probe alignment relative
to the detector, the vertical electric field uniformity inside the
detector, and most significantly, the 1 cm total device length in
transmission mode. This length degrades resolution through beam
divergence, refraction, and scattering, effectively increasing the
minimum resolvable feature size. Combining these contributions in
quadrature, the overall spatial resolution is conservatively estimated
to be in the range of 10--15 $\mu$m, which is smaller than the
minimum measured widths of the electron clouds
($\sigma \approx 20$ $\mu$m, corresponding to 
a full width at half maximum [FWHM] of $\approx 45$ $\mu$m). The sample has
been conditioned each time before a given experimental sequence by
keeping it under the selected applied voltage and pump beam power for
15--20 min. This allowed it to reach a quasi-stationary condition
before starting the CDPI acquisition (Fig.~S2).

In the sequence, a set of images is registered upon changing the delay
between pump and probe by fixed time intervals. Usually, the time step
is either 0.5 ns or 1.0 ns in the short-time range case (up to a total
200 ns) or either 10 ns or 20 ns in the long-time range (up to 2
$\mu$s). Before and after the CDPI acquisitions with the pulsed probe,
a continuous light-emitting diode (LED) filtered at 980 nm and impinging on the sample using a
beam-splitter, was instead used as check and reference probe beam. For
the whole protocol procedure see Fig.~S2. The experimental setup and
the sequence of different instrument operations have been coordinated
by means of a custom program in the NI Labview graphical language. The
external current transients have been taken under similar conditions
adopted for the Pockels experiments, with corresponding applied
voltages and pump beam power. The sample has been connected to an
amplifier (Mini-Circuit, bandwidth 1 GHz) and the output signal fed to
a 1 GHz oscilloscope (Lecroy), triggered by the pump laser output. The
detector voltage bias is applied on the same electrode, isolated from
the amplifier input by a capacitor.\\

\subsection*{Electro-optic imaging}
The intensity transmitted throughout cross-polarizers is zero, in the
case of the probe passing unaffected through the unbiased sample. This
changes in the presence of internal birefringence (intrinsic or induced
by an applied electric field), according to a modulation formula
\cite{yariv_optical_1991}, and the scheme can convert any phase
difference (here between the $x$ and $y$ linear polarizations) into an
intensity signal. With the present orientation of the polarizers and of
sample crystal planes, the intensity is sensitive only to the $y$
component of the externally applied electric field. Such $E_y$ is
effectively the only present field component due to full-surface planar
electrode geometry and wide pumping beam, inducing a vertical gradient
of the electric potential inside the device. The raw $P(x,y,t)$ images
depend on the $E_y(x,y,t)$ electric field distribution, and they can be
normalized with respect to $P_{\mathrm{para}}(x,y)$ as:
\begin{equation}
T(x,y,t)
  = \frac{P(x,y,t)}{P_{\mathrm{para}}(x,y)}
  = \sin^{2}\!\left[\frac{\pi}{2}\frac{E_y(x,y,t)}{E_0}\right].
\label{eq:M1}
\end{equation}
Here $P_{\mathrm{para}}(x,y)$ is the image at the maximum transmitted
intensity of the probe, taken in the parallel-polarizer configuration
(both aligned at 45$^{\circ}$ and in the absence of any induced
birefringence, i.e., with no applied voltage). The use of normalization
in Eq.~\eqref{eq:M1} is crucial in the analysis because beside enabling
quantitative evaluation, it can partially account for reflection loss,
non-homogeneous distribution of impinging intensity and other device
imaging defects. The constant $E_0$ is given by:
\begin{equation}
E_0 = \frac{\lambda}{\sqrt{3}\,n_0^{3}\,r_{41}\,L},
\label{eq:M2}
\end{equation}
where $n_0$ is the field-free refractive index in CdTe, $r_{41}$ its
linear electro-optic coefficient, $L = 10$ mm is the physical path
length through the crystal, i.e., the detector $z$-depth, and $\lambda$
the wavelength of the incident probe light. Considering $n_0 = 2.8$ and
$r_{41} = 5.5$ pm/V \cite{bylsma_photorefractive_1987} results in
$E_0 = 4.68$ kV cm$^{-1}$. The meaning of $E_0$ is that it represents
the local electric field value that allows a complete polarization
inversion of the crossing probe beam, hence the Pockels image bears
maxima (equal to $P_{\mathrm{para}}$) and minima (zero) intensity
values at the transverse $xy$ positions where the electric field is
equal to odd and even multiples of $E_0$, respectively. When in the
presence of local electric-field values greater than $E_0$, multiple
fringes appear in the Pockels $T(x,y)$ images and an unwrapping-phase
reconstruction is needed to achieve the $E(x,y)$ profile. In the
present work electric fields do not exceed $E_0$, and a simple
inversion of the square sinusoidal dependence in Eq.~\eqref{eq:M1}
allows to obtain the $E_y(x,y)$ image from the normalized EO images.
From the $E_y(x,y)$ images, average $E_y(y)$ profiles are obtained and
smoothed.\\

\subsection*{Carrier transport in CDPI experiments}
The Gauss's law, at any given time, relates the electric field to the
mobile and fixed charges inside the device
\cite{weisz_electrode-limited_1968,batra_photocurrents_1970}:
\begin{equation}
\frac{\partial E(y,t)}{\partial y}
  = \frac{q}{\varepsilon}\left(n_t(y,t) + \delta n(y,t)\right).
\label{eq:M3}
\end{equation}
Here $\delta n(y,t)$ represents the free electron cloud density and
$n_t(t,y)$ the fixed space charge density. In general, electron
trapping time in CdTe is about 1 $\mu$s, which is greater than the
transit time under operative electric fields
($> 600$ V cm$^{-1}$ in Pt/CdTe/Pt detectors) and thus not effective.
Under these conditions, $n_t(t,y)$ remains stationary during the cloud
drifting and equals the values prior to photogeneration at $t = 0$,
$n_t(y,t) \approx n_t(y,t=0)$. This means that there is a stationary
electric field distribution on which a perturbation is transiently
superimposed due to the flowing charge. Depending on the ratio between
the free and fixed charges, the perturbation can be small or very
important, up to becoming comparable or greater than the background.
Under high optical excitation levels, electron trapping is enhanced by
specific shallow levels thus directly affecting both charge terms in
Eq.~\eqref{eq:1}, namely the mobile carrier density $\delta n(y,t)$,
and the fixed charge density $n_t(t,y)$.

In the ideal case of uniform electric field and negligible trapping,
because of Gauss's law, the electric field is reduced in the regions
behind its position and enhanced in front of it by a total difference
of
\begin{equation}
\Delta E = \frac{q}{\varepsilon}N_S,
\label{eq:M4}
\end{equation}
where
\begin{equation}
N_S = \int \delta n(y,t)\,\mathrm{d}y
\label{eq:M5}
\end{equation}
is the areal charge density.

By assuming a constant carrier mobility $\mu$ and that the diffusion
component is negligible, the conduction current can be written as only
due to the drift term
\begin{equation}
J_{\mathrm{con}}(y,t) = q\,\mu\,E(y,t)\,\delta n(y,t)
\label{eq:M6}
\end{equation}
at any given point (and time) inside the device. The diffusion term can
be ignored in the present range of electric fields, and we further
verified this by means of numerical simulations.

The total current $J(t)$ is due to the sum of two terms, the conduction
current $J_{\mathrm{con}}(y,t)$ and the displacement current $J_{\mathrm{dis}}(y,t)$, and is
given by:
\begin{equation}
J(t)
  = q\,\mu\,E(y,t)\,\delta n(y,t)
  + \varepsilon\,\partial E(y,t)/\partial t.
\label{eq:M7}
\end{equation}
By assuming a negligible trapping on the scale of transit time, we can
write
$n_t(y,t) = \frac{\varepsilon}{q}\frac{\partial E(y,t=0)}{\partial y}$ in
Eq.~\eqref{eq:M3} and thus express the total current in
Eq.~\eqref{eq:M7} only in terms of the electric field:
\begin{equation}
J(t)
  = \mu\,\varepsilon\,E(y,t)
    \left[
      \frac{\partial E(y,t)}{\partial y}
      - \frac{\partial E(y,t=0)}{\partial y}
    \right] \newline
    + \varepsilon\,\partial E(y,t)/\partial t.
\label{eq:M8}
\end{equation}
Operatively, the spatiotemporal distribution of the carrier cloud can
be calculated either directly, from the spatial derivative of the
electric field (Eq.~\eqref{eq:M3}) or, indirectly, from the
complementary displacement current (Eq.~\eqref{eq:M8}), according to:
\begin{equation}
\delta n_{\mathrm{displ}}(y,t)
  = -\,\frac{\varepsilon\,\partial E(y,t)/\partial t - J(t)}
           {q\,\mu\,E(y,t)},
\label{eq:M9}
\end{equation}
where the $\delta n_{\mathrm{displ}}$ notation makes explicit the
derivation of this quantity based on the displacement current.

The displacement current method represents a significant technical
advantage in terms of data analysis, as it allows us to infer the
carrier cloud from the temporal derivative of the electric field, far
less noisy than the spatial derivative. The total $J(t)$ contribution
in Eq.~\eqref{eq:M9} is in general small with respect to the
displacement term in the region of the electron cloud. Therefore, when
focusing on the (non-quantitative) spatial distribution of the electron
cloud, the total current can be reasonably neglected, and:
\begin{equation}
\delta n_{\mathrm{displ}}(y,t)
  \sim -\,\frac{\varepsilon\,\partial E(y,t)/\partial t}{q\,\mu\,E(y,t)}.
\label{eq:M10}
\end{equation}
A further simplification in case of relatively uniform electric field
(or very thin electron cloud) would be to neglect the field spatial
dependence, which results in a direct correspondence between the shapes
of $\delta n_{\mathrm{displ}}(y,t)$ and of the displacement current.
This is suitable in the case of low optical photogeneration.

According to Eq.~\eqref{eq:M8}, the CDPI technique offers, in principle,
the remarkable opportunity to extract the transient current flowing in
the external circuit from the sum, at any arbitrary point in the
detector, of the conduction current, related to the spatial derivative
of the electric field (and the electric field itself), and the
displacement current, related to the temporal derivative of the
electric field.

As shown in literature \cite{weisz_electrode-limited_1968,batra_photocurrents_1970},
a more direct relationship current vs electric field emerges from the
spatial integration of Eq.~\eqref{eq:M8}, taking into account that
$J(t)$ is independent of $y$, that the integral of the displacement
current term is zero, and that the following assumptions hold:\\

$\medblackcircle$~Uniform electric field in absence of carrier excess\\

~~~$\left( \frac{\partial E(y,t=0)}{\partial y} = 0 \right)$;\\
 
$\medblackcircle$~Trapping is neglected.\\

Another implicit assumption is that only one type of carrier is considered.
This considered, we obtain:
\begin{equation}
J_{\mathrm{approx}}(t)
  = \frac{\varepsilon\,\mu}{2L}
    \left[E^{2}(L,t) - E^{2}(0,t)\right],
\label{eq:M11}
\end{equation}
where, as anticipated, the dependence on the electric field rather than
on its derivative and only in two fixed points at the electrodes, is very
useful when feeding this into the CDPI technique. With respect to the
above-mentioned approximations, we can only partially circumvent the
first one: the electric field is not uniform due to the presence of a
fixed space charge. We add a constant baseline to Eq.~\eqref{eq:M11} to
ensure $J(t=0) = 0$. It should be noted though that when integrating
Eq.~\eqref{eq:M8} the presence of a space charge affects the entire
transient.\\

\subsection*{Numerical methods}
1D numerical simulations were performed using the semiconductor device
simulator ``Sentaurus'', part of the Technology CAD software package
provided by Synopsys, Inc.~\cite{noauthor_sentaurus_nodate}. The basic elements
of a model that are able to describe the CdTe device under applied bias
and light in quasi-stationary conditions have been previously presented
\cite{cola_electric-field_2023} and are reported here in full
(Table~S2). In this work, the model has been extended to reproduce the
experimental transients on the electron transit timescale, and, to this
end, we have included the presence of shallow traps, according to the
different experimental conditions reported in the text. We simulated
the whole experimental flow in the time domain, starting from applying
bias voltage under dark (according to the protocol in Fig.~S2) and by
considering optical pulses 1 ns long at a frequency of 25 kHz when
illumination is active. Simulations are performed using the
Backward-Euler method using an adaptive time step for ensuring
convergence and controlling numerical errors. Data have been recorded at
0.5 ns or 1 ns interval, matching the experimental resolution to
facilitate a direct comparison.

\section*{Resource availability}

\subsection*{Lead contact}
Requests for further information and resources should be directed to
and will be fulfilled by the lead contact, Adriano Cola
(adriano.cola@cnr.it).

\subsection*{Materials availability}
This study did not generate new materials. The original devices were
provided by Acrorad as stated in the Methods.

\subsection*{Data and code availability} 
Original data used in this work as well as any additional information
required to reanalyze the data reported in this paper are available
from the corresponding authors upon reasonable request.

\section*{Acknowledgements}
A.C.~is grateful to Massimiliano Zanichelli for the initial motivation
and inspiration. L.D.~wishes to thank Francesco Michelotti for the
initial inspiration in experimental physics, optical nonlinearities and
the Teng-Man technique. A.C.~and L.D.~thank Paolo Cazzato for
technical support, and Marc Currie and Carlo Giansante for scientific
advice. This research was partially funded by the PNRR MUR project
PE0000023-NQSTI, financed by the European Union - Next Generation EU.
We acknowledge the ELPHO laboratory at CNR NANOTEC where all the
experiments were performed.

\section*{Author contributions} 
A.C.~and L.D.~conceived the idea and experimental procedure.
A.C.~designed and implemented the CDPI setup. L.D.~wrote the software for
experiment control and analysis. A.C.~conducted the experiments.
I.F.~helped with the setup. A.C.~and L.D.~analysed the results.
A.V.~conducted numerical simulations. V.D.~provided important expertise
on the electro-optic effect. I.F.~contributed to funding acquisition.
A.C.~and L.D.~wrote the manuscript. All authors contributed to the
discussion and interpretation of the data and revised the manuscript.

\section*{Declaration of interests}
The authors declare no competing interests.\\

\section*{Declaration of generative AI and AI-assisted technologies in the manuscript preparation process}
During the preparation of this work, the authors used ChatGPT for
language editing and stylistic refinement. After using this tool, the
authors reviewed and edited the content as needed and take full
responsibility for the content of the publication. DeepSeek and ChatGPT were also used to assist 
in formatting the \LaTeX{} script of the arXiv version.\\

\section*{Supplemental Information}
The \textbf{\href{https://arxiv.org/src/2607.19157v2/anc/CDPI_Supplemental_Info.pdf}{Supplemental Information}},
 including a PDF file with three supplementary
figures and two supplementary tables, and four supplementary videos
(Videos S1--S4), is available as ancillary files in the arXiv version
of this article and on the published version at the journal page \href{https://doi.org/10.1016/j.matlit.2026.100141}{Matter \& Light 1, 100141, 2026}.\\

\subsection*{\href{https://arxiv.org/src/2607.19157v2/anc/Video-S1.mp4}{Video S1. Temporal evolution of the transmittance difference maps}}
Temporal evolution of the transmittance difference maps
$T(x,y,t+\mathrm{d}t)-T(x,y,t)$ similarly to Figure~1B, over an
80~ns time range and with a step of 1~ns (the interval for differences
is taken as $\mathrm{d}t = 2$~ns). The moving strip is representative
of the drifting electron cloud. The three cases refer to increasing
optical irradiation powers, left to right, corresponding, respectively,
to $P_4$, $P_6$, and $P_8$ discussed in Figure~3. The applied voltage
is 200~V. The color scale is set between $-0.02$ and $+0.02$ among all
the maps to allow visual comparison and to optimize the strip contrast.
The two horizontal solid black lines represent the cathode (top) and
anode (bottom) electrodes.\\

\subsection*{\href{https://arxiv.org/src/2607.19157v2/anc/Video-S2.mp4}{Video S2. Temporal evolution of electric field profile and of its relative variation}}
Electric field profile (black line) and its relative variation
$[E(y,t) - E(y,t=0)] / E(y,t=0)$ (orange line) as in Figure~2C; time
evolution over 60~ns (time step 0.5~ns). The $y$-profiles (from cathode
to anode) have been calculated by averaging on the $x$ direction, along
which the field is uniform. Data refer to the low optical irradiation
case (same as Figure~2C and $P_4$ in Figure~3).\\

\subsection*{\href{https://arxiv.org/src/2607.19157v2/anc/Video-S3.mp4}{Video S3. Temporal evolution of displacement current profiles at different applied voltages}}
Temporal evolution of displacement currents moving from cathode to
anode over 125~ns (time step 1~ns) at different applied voltages:
50~V (green), 100~V (yellow), 150~V (red), and 200~V (blue) in the
case of low optical irradiation (Figure~2D and $P_4$ in Figure~3).
In this low-power case, the displacement current peak is Gaussian and
representative of the electron cloud position and shape.\\

\subsection*{\href{https://arxiv.org/src/2607.19157v2/anc/Video-S4.mp4}{Video S4. Temporal evolution of carrier cloud profiles under different optical irradiation levels}}
Temporal evolution of carrier cloud profiles over a 72~ns time range
(time step 1~ns) under different optical irradiation levels as in
Figure~3F. The excitation optical density is increasing from low to
high perturbation as $P_2$ (brown), $P_4$ (yellow), $P_6$ (blue),
$P_8$ (red). The applied bias is 200~V. By increasing the optical
irradiation level, the carrier clouds reshape from purely Gaussian to
bi-Gaussian with more extended tail than front.\\


%

\end{document}